\documentclass{article}

\usepackage{PRIMEarxiv}

\usepackage[utf8]{inputenc} 
\usepackage[T1]{fontenc}    
\usepackage[hidelinks]{hyperref}
\usepackage{url}            
\usepackage{booktabs}       
\usepackage{amsfonts}       
\usepackage{nicefrac}       
\usepackage{microtype}      
\usepackage{lipsum}
\usepackage{fancyhdr}       
\usepackage{graphicx}       
\graphicspath{{media/}}     

\usepackage{listings}
\usepackage{amsmath,amssymb,amsfonts}
\usepackage{algorithmic}
\usepackage{textcomp}

\usepackage{float}
\usepackage{subfiles}
\usepackage{subcaption}
\usepackage{rotating}
\usepackage{adjustbox}
\usepackage[table,xcdraw]{xcolor}
\usepackage{lscape}
\usepackage{tabularx}
\usepackage{tcolorbox}

\title{Automated Generation of Accurate Privacy Captions from Android Source Code using Large Language Models}

\author{
  Vijayanta Jain, \\
  University of Maine \\
  Orono, ME\\
  \texttt{vijayanta.jain@maine.edu} \\
  \And
  Sepideh Ghanavati\\
  University of Maine \\
  Orono, ME\\
  \texttt{sepideh.ghanavati@maine.edu} \\
  \And
  Sai Teja Peddinti \\
  Google Inc. \\
  Mountain View, CA\\
  \texttt{psaiteja@google.com} \\
  \And
  Collin McMillan \\
  University of Notre Dame \\
  Notre Dame, IN \\
  cmc@nd.edu \\ }

\begin{document}
\maketitle

\begin{abstract}
Privacy captions are short sentences that succinctly describe \textit{what} personal information is used, \textit{how} it is used, and \textit{why}, within an app. These captions can be utilized in various notice formats, such as privacy policies, app rationales, and app store descriptions. However, inaccurate captions may mislead users and expose developers to regulatory fines. Existing approaches to generating privacy notices or just privacy captions include using questionnaires, templates, static analysis, or machine learning. However, these approaches either rely heavily on developers' inputs and thus strain their efforts, use limited source code context, leading to the incomplete capture of app privacy behaviors, or depend on potentially inaccurate privacy policies as a source for creating notices. In this work, we address these limitations by developing \textbf{P}rivacy \textbf{Cap}tion \textbf{Gen}erator (PCapGen), an approach that - i) automatically identifies and extracts large and precise source code context that implements privacy behaviors in an app, ii) uses a Large Language Model (LLM) to describe coarse- and fine-grained privacy behaviors, and iii) generates accurate, concise, and complete privacy captions to describe the privacy behaviors of the app. Our evaluation shows PCapGen generates concise, complete, and accurate privacy captions as compared to the baseline approach. Furthermore, privacy experts choose PCapGen captions at least 71\% of the time, whereas LLMs-as-judge prefer PCapGen captions at least 76\% of the time, indicating strong performance of our approach. 
\end{abstract}

\section{Introduction}

App privacy notices are important artifacts that describe \textit{what} types of personal information are being used in an app, \textit{how} they are used, and \textit{why} (i.e., the privacy behaviors of the app). We define \textit{privacy captions} as natural language sentences that describe privacy behaviors of an app. A sample privacy caption might look like: ``This app collects your location data, device identifiers, and network information to deliver personalized advertisements.''.  Almost all formats of privacy notices, such as privacy policies, rationales, in-app notices  (i.e., disclosures shown at runtime), and app descriptions, use privacy captions. These captions are crucial for all stakeholders involved, including app developers, users, app marketplaces, and regulators. Users read these captions to understand apps' privacy behaviors to make informed privacy decisions. Regulators and marketplaces use privacy captions to ensure that the privacy behaviors described in the notice align with the app's actual behavior and comply with regulations \cite{CCPA, GDPR} and policies \cite{AppStoreLabels, PlayLabels}.

The challenge is that the responsibility of creating accurate captions and ensuring their consistency across all notice formats falls on developers. This especially strains the efforts of those working alone or in small teams, and cannot hire privacy experts \cite{li2022Understanding, Khandelwal2024}. Consequently, these developers may use other methods to create captions or privacy notices. For example, they may use online policy generators to create privacy policies \cite{prybylo2024evaluating} or copy and paste from similar apps \cite{panita}. In both cases, the resultant policy may not describe all the app's privacy behaviors  \cite{jain2023atlas}. Developers may also skip creating permissions rationales or create incorrect ones \cite{liulargescale2018}, as they may lack an understanding of the privacy behavior of their app when using third-party libraries \cite{jain2023towards}. Lastly, developers reuse the same rationales in app descriptions~\cite {liu2018large}, and thus the descriptions end up being inaccurate too~\cite{gorla2014checking}. All these practices suggest that the privacy captions often do not match the privacy behaviors of the app \cite{jain2023atlas, liu2018large, gorla2014checking}, and they also vary across different notice formats. Inconsistency between the privacy caption and the privacy behaviors violates privacy regulations \cite{gdpr_transparency, ccpa_notice_requirement} and app store policies, and may result in monetary fines~\cite{edpb2023} or removal of such apps from marketplaces~\cite{play_store_enforcement, app_store_enforcement}.
Several efforts have been directed to help developers create privacy notices~\cite{li2021honeysuckle, zimmeck2021privacyflash, liao20243, yu2016toward, gardnerhelping2022, jain2023towards, jain2022pact}. Some approaches combine static analysis of the app's code with questionnaires to create privacy policies \cite{zimmeck2021privacyflash}, or with developers' manual source code annotations describing privacy behaviors to create in-app notices \cite{li2021honeysuckle}. There are also machine learning based approaches to generate privacy captions \cite{jain2021prigen} and privacy labels \cite{jain2023towards}. Recent works have also used Large Language Models (LLMs) to create in-app notices~\cite{pan2024new} and privacy labels~\cite{pan2023toward} by analyzing the apps' UI and privacy policies. 

While all these approaches are helpful to generate a specific privacy notice format, there are several limitations. First, most of these approaches rely on significant efforts from developers to create these notices \cite{zimmeck2021privacyflash, gardnerhelping2022, li2021honeysuckle}, such as manually annotating their source code with privacy behavior descriptions \cite{li2021honeysuckle}, even when developers often do not completely understand the privacy behaviors of their app \cite{hadar2018privacy}. Second, automated approaches either use limited context or potentially inaccurate context to create privacy notices. For example, Jain et al. \cite{jain2022pact, jain2023towards} only use up to three methods in the call graph to create privacy labels. Pan et al. combine static analysis of an app's UI with privacy policy analysis to create privacy labels \cite{pan2023toward} or in-app contextual notice \cite{pan2024new}. Since research has shown that privacy policies often incorrectly describe the privacy behaviors \cite{zimmeck2019maps}, the resultant privacy labels and in-app notices may also be inaccurate \cite{jain2023atlas}. Third, these approaches primarily focus on creating a specific privacy notice format, such as a privacy policy or an in-app privacy notice. While helpful, it requires developers' effort to extract the privacy caption from these notices and transform them into acceptable language for other formats. For example, sentences in privacy policies are filled with legal jargon, which can be too complex for a permission rationale. On the other hand, privacy captions are natural language sentences that can be used in all notice formats.

In this paper, we address the limitations of previous works by proposing a novel framework called \textbf{P}rivacy \textbf{Cap}tion \textbf{Gen}erator (PCapGen). PCapGen contains three main components to automatically create accurate privacy captions: i) \textit{Identifier} uses heuristics-based taint and static analysis to automatically identify taint paths in an app that describe the flow of personal information. ii) \textit{Extractor} uses static analysis to extract the source code of all the classes and methods involved in the taint paths and uses Large Language Models (LLMs) to identify individual statements in the source code that use personal information. iii) \textit{Generator} uses LLMs with In-Context Learning (ICL)~\cite{brown2020language} to comprehend the privacy behaviors implemented in the source code and generates accurate privacy captions. The novelty of PCapGen lies in developing a heuristics-based taint analysis that identifies taint paths between new source API methods and unknown sink methods, and in developing a hybrid approach that complements static analysis with LLMs to automatically construct a large source code context that spans multiple statements in a method, multiple methods in a class file, and multiple class files, giving us both, a coarse-grained \textit{and} a fine-grained view of \textit{what} personal information is used, \textit{how} it is used, and \textit{why}. This large source code context helps the \textit{Generator} create concise, complete, and accurate privacy captions that can be used in various privacy notice formats. 

We demonstrate the robustness of PCapGen by using three different LLMs: GPT 4 (gpt-4-0613) \cite{achiam2023gpt}, Claude Opus 4 (claude-opus-4-20250514) \cite{claudeopus4}, and DeepSeek (deepseek-r1-distill-llama-70b)\cite{liu2024deepseek}, and evaluating them quantitatively and qualitatively with privacy experts, LLMs-as-judge, and automated metrics. We evaluate our approach and the captions generated, in comparison to developer-curated baseline captions, by answering the following research questions:

\begin{enumerate}
     \item \textbf{RQ 1.} Which  PCapGen configuration generates better privacy captions as per different quality criteria?
    \item \textbf{RQ 2.} How similar are the PCapGen privacy captions compared to the baseline privacy captions?
   \item \textbf{RQ 3.} Are (best configuration) PCapGen privacy captions better than baseline privacy captions on the quality criteria according to LLM-as-judge models?
   \item \textbf{RQ 4.} Are (best configuration) PCapGen privacy captions better than baseline privacy captions on the quality criteria according to privacy experts?
\end{enumerate}

Our results show that PCapGen\_Claude performs the best among the three model configurations on all quality criteria - accuracy, conciseness, and completeness (Section \ref{subsec:config_comparison}). All three PCapGen configurations generate captions that are semantically very similar to the baseline captions. LLMs-as-judge rate PCapGen\_Claude captions higher than the baseline captions on all quality criteria (Section \ref{subsec:llm-results}) and prefer PCapGen\_Claude captions at least 76\% of the time over baseline captions. Lastly, while we do not find a statistically significant difference between the baseline and PCapGen captions based on the privacy experts' quality ratings, experts prefer PCapGen\_Claude captions 71\% of the time over baseline captions. These results indicate the superior performance of the PCapGen framework in automatically generating high quality captions that are as good as or better than developer curated captions.


The following are the primary contributions of this work:

\begin{enumerate}
    \item \textbf{Novel Approach}: We provide a novel framework that complements traditional static and taint analysis with LLMs' ICL to identify coarse and fine-grained privacy behaviors implemented in source code and generate accurate privacy captions.  
    \item \textbf{Extended Taint Analyses}: We develop a new heuristics-based approach for taint analysis that increases the scalability of traditional approach by extracting taint paths of new sources and unknown sinks, and the associated source code of the classes, methods, and their statements.
    \item \textbf{Comprehensive Expert and LLM-as-Judge Evaluations}: We demonstrate the efficacy of PCapGen by evaluating it with privacy experts and LLMs-as-judge models, providing a strong evaluation framework that can be used by the research community to evaluate privacy artifacts. 
    \item \textbf{Dataset and Tools}: We share the high-quality dataset that contains developer-annotated Android source code snippets with privacy captions and scripts used to generate and evaluate them and our custom annotation tool used for our survey in this paper.
\end{enumerate}


\label{sec:introduction}

\section{Related Work}

We discuss how our approach relates to prior works in this section.  



\begin{table*}[htbp]
\caption{Summary of related approaches to create privacy notices}
\label{tab:related-works}
\footnotesize
\begin{tabularx}{\textwidth}{|p{2.7cm}|X|p{3.2cm}|p{3.4cm}|c|c|} 
\hline
\textbf{Work} & \textbf{Input Context} & \textbf{Privacy Notice Format} & \textbf{Developer's Input} & \textbf{LLM} & \textbf{Automated} \\ \hline
PrivacyFlash Pro \cite{zimmeck2021privacyflash} & API Calls, Import Statements, Questionnaires & Privacy Policy & Questionnaires, Permission Rationales & $\times$ & Partial \\ \hline
PrivacyLabel Wiz \cite{gardnerhelping2022} & API Calls, Import Statements, Questionnaires & Privacy Labels & Questionnaires, Permission Rationales & $\times$ & Partial \\ \hline
DescribeCTX & Call Graph, App Description, Privacy Policies & Privacy Captions & None & $\times$ & $\checkmark$ \\ \hline
PriGen \cite{jain2021prigen} & Single Method & Privacy Caption & None & $\times$ & $\checkmark$ \\ \hline
Jain et al. \cite{jain2022pact, jain2023towards} & Three methods in call graph & Privacy Labels & None & $\times$ & $\checkmark$ \\ \hline
HoneySuckle \cite{li2021honeysuckle} & Developer written descriptions & In-App Notice & Annotation \& rationales & $\times$ & Partial \\ \hline
Matcha \cite{limatcha2024} & Source code + SDK analysis & Privacy Labels & Annotation \& verification & $\times$ & Partial \\ \hline
SeePrivacy \cite{pan2024new, panseeprivacy2023} & GUI \& privacy policies & In-App Notice & None & $\checkmark$ & $\checkmark$ \\ \hline
Pan et al. \cite{pan2023toward} & Privacy policies & Privacy Labels & None & $\checkmark$ & $\checkmark$ \\ \hline
\textbf{PCapGen} & Taint Paths, Classes, & Privacy Captions & None & $\checkmark$ & $\checkmark$ \\ 
\textbf{(This Work)} & Methods, and Statements & & & & \\
\hline
\end{tabularx}
\end{table*}

\textbf{Discrepancies between Notice and Behaviors.}
Several studies have examined discrepancies between a privacy notice and the actual privacy behaviors of an app~\cite{jain2023atlas, zimmeck2019maps, liu2024ihunter, alihonesty2024, buiconsistency2021, liu2018large, slavin2016pvdetector, okoyomon2019ridiculousness, maitra2018privacy, xiao2022lalaine, ali2023honesty, alomar2025effect}. These efforts vary based on the type of notice, which includes privacy policies \cite{zimmeck2019maps, maitra2018privacy}, app descriptions \cite{gorla2014checking, qu2014autocog, pandita2013whyper}, permission rationales \cite{liu2018large}, and privacy labels \cite{xiao2022lalaine, jain2023atlas}. Other works have investigated the inconsistency between privacy policies and privacy labels \cite{ali2023honesty, khandelwaloverview2023, jain2023atlas}, revealing that different notice formats often amplify inconsistencies. 
Most prior research primarily develop approaches to identify these discrepancies at scale and are intended for regulators and app marketplaces rather than developers. 
While such approaches expose the extent of the problem at a large scale, they provide limited support in mitigating or resolving the discrepancies. Our approach, however, aims to resolve these inconsistencies by directly generating traceable privacy captions.



\textbf{Creating Privacy Notices.}
Several approaches create different privacy notice formats, such as privacy policies \cite{zimmeck2021privacyflash, pan2024new, yu2016toward}, privacy labels \cite{pan2023toward, gardnerhelping2022, limatcha2024a, jain2022pact, jain2023towards}, privacy captions \cite{jain2021prigen, yang2022describectx}, and in-app notices \cite{li2021honeysuckle, taoprivacy2025, panseeprivacy2023}. Table \ref{tab:related-works} summarizes prior work in compare to our PCapGen. While helpful, there are four primary limitations. 

\emph{First}, most approaches use very limited source code context when creating privacy notices \cite{zimmeck2021privacyflash, gardnerhelping2022, jain2021prigen, jain2022pact, jain2023towards}. PrivacyFlash Pro \cite{zimmeck2021privacyflash} only uses the API calls and library imports as the source code context, along with developer-written permission rationales and a questionnaire to create privacy policies. However, it does not include other methods that might use the information returned by the API call; hence, creating accurate privacy policies is based solely on developers' understanding of the privacy behaviors. Privacy Label Wiz \cite{gardnerhelping2022} utilizes the same source code context and a developer questionnaire to create privacy labels. Jain et al. \cite{jain2022pact, jain2023towards} extend this source code context to use up to three methods in the call graph to automatically create privacy labels. However, this context also might not capture the privacy behaviors implemented beyond the three methods.  Lastly, DescribeCTX \cite{yang2022describectx} analyzes call graphs, app descriptions, and privacy policy statement examples to generate privacy captions. However, it only analyzes the method's name in the call graph and not the source code. 
\emph{Second}, most studies rely on the developer's input to create privacy notices~\cite{zimmeck2021privacyflash, gardnerhelping2022, li2021honeysuckle, limatcha2024}, which require them to repeatedly put in this extra effort every time they need to update their privacy notices as their app evolves. This increases the burden on developers, especially those working alone or in small teams. HoneySuckle \cite{li2021honeysuckle} creates disclosures shown during runtime but requires developers to annotate the source and sink methods and provide an explanation for why the information is used in these methods. While Matcha \cite{limatcha2024} automates the analysis of identifying the source and sink methods to create privacy labels, it requires developers to verify the annotation. It also primarily focuses on identifying sinks where the personal information is saved off-device, which could exclude scenarios where the data is processed within the app.
\emph{Third}, some approaches use LLMs to create privacy notices, but use potentially inaccurate context to create them \cite{pan2023toward, pan2024new, panseeprivacy2023}. SeePrivacy \cite{pan2024new, panseeprivacy2023} analyzes the GUI context of Android apps and displays relevant privacy policy text during runtime. 
Pan et al. \cite{pan2023toward} use LLMs to transform privacy policies into privacy labels for the App Store and Play Store. While these approaches automate the process of creating privacy notices, they use the privacy policy as context, which, as discussed previously, has been shown to be incorrect \cite{zimmeck2019maps}. 
\emph{Lastly}, these approaches primarily focus on creating a specific privacy notice format, such as a privacy policy \cite{zimmeck2021privacyflash} or an in-app privacy notice \cite{panseeprivacy2023}, which still requires developers' effort to extract the privacy caption from them and transform them into acceptable language for other formats. For example, sentences in privacy policies are filled with legal jargon, which can be too complex for a permission rationale. This process can be strenuous on developers lacking privacy expertise. On the other hand, privacy captions are natural language sentences that can be used in all notice formats.

\textit{PCapGen} proposes to address these limitations by using the complete source code context from taint paths to identify privacy behaviors and automatically generate privacy captions without any developers' input, which could be used in several different privacy notice formats. 

\begin{figure*}[thbp]
    \centering
    \includegraphics[width=0.9\textwidth]{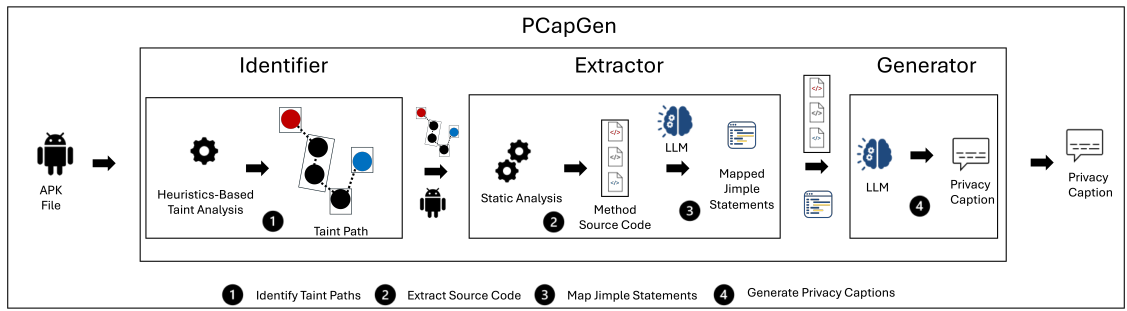}
   \caption{Overview of our PCapGen describing the key components and steps to generate privacy captions.}
      \label{fig:overview}
\end{figure*}

\textbf{Generating Code Documentation.}
Our approach is also inspired by efforts in software engineering to document code. Documentation can take the form of generating comments or JavaDocs of individual methods \cite{alon2018code2seq, hu2018deep, haque2020improved, chen2019neural, geng2024large, sun2023automatic, sun2024source, su2024distilled, wu2025can} or generating commit messages of the diff in the source code \cite{jiang2017automatically, loyola2017neural, fan2025exploring, dong2022fira, li2024only, eliseeva2023commit}. While earlier efforts focused on developing dedicated deep learning models to document source code \cite{alon2018code2seq, hu2018deep, haque2020improved, chen2019neural, jiang2017automatically, loyola2017neural, dong2022fira}, most recent works use LLMs for this process \cite{geng2024large, sun2023automatic, sun2024source, su2024distilled, wu2025can, eliseeva2023commit, li2024only, fan2025exploring}. Although our goal resembles code comment generation efforts to briefly describe the source code, it is more similar to commit message generation because it also analyzes a large source code context that spans multiple files, methods, and statements.

\label{sec:bg-and-related-work}

\section{Proposed PCapGen Approach}

The goal of PCapGen is to \textit{identify} the privacy behaviors implemented in an app, \textit{extract} the source code implementing them, and \textit{generate} privacy captions to describe them. We develop a corresponding component in PCapGen framework (Figure~\ref{fig:overview}) to complete each of these tasks. Identifier uses heuristics-based taint analysis to identify taint paths that describe the flow of personal information within an app. Extractor uses static analysis to extract the source code of classes and methods from  taint paths, and LLMs to map individual statements using personal information in this source code. Lastly, Generator uses an LLM to describe the privacy behaviors in the extracted source code context to generate concise, complete, and accurate privacy captions. 


\subsection{Identifier: Identifying Taint Paths}
\label{subsec:identifier}

Identifier aims to identify the path of personal information, i.e., the classes, methods, and statements through which personal information flows within an app. This flow or path describes the apps' privacy behavior, i.e., \textit{what} personal information is used, \textit{how} it is used, and \textit{why}.  
The component uses taint analysis to identify taint paths that describe the flow of information. We provide a brief overview of the taint analysis process, followed by our heuristics-based taint analysis approach for extracting taint paths. We use FlowDroid for this task as it has been used extensively in research \cite{sun2024llm4vuln, wang2023taintmini, ami2024false}. 

\begin{figure}
    \centering
    \includegraphics[width=0.999\columnwidth]{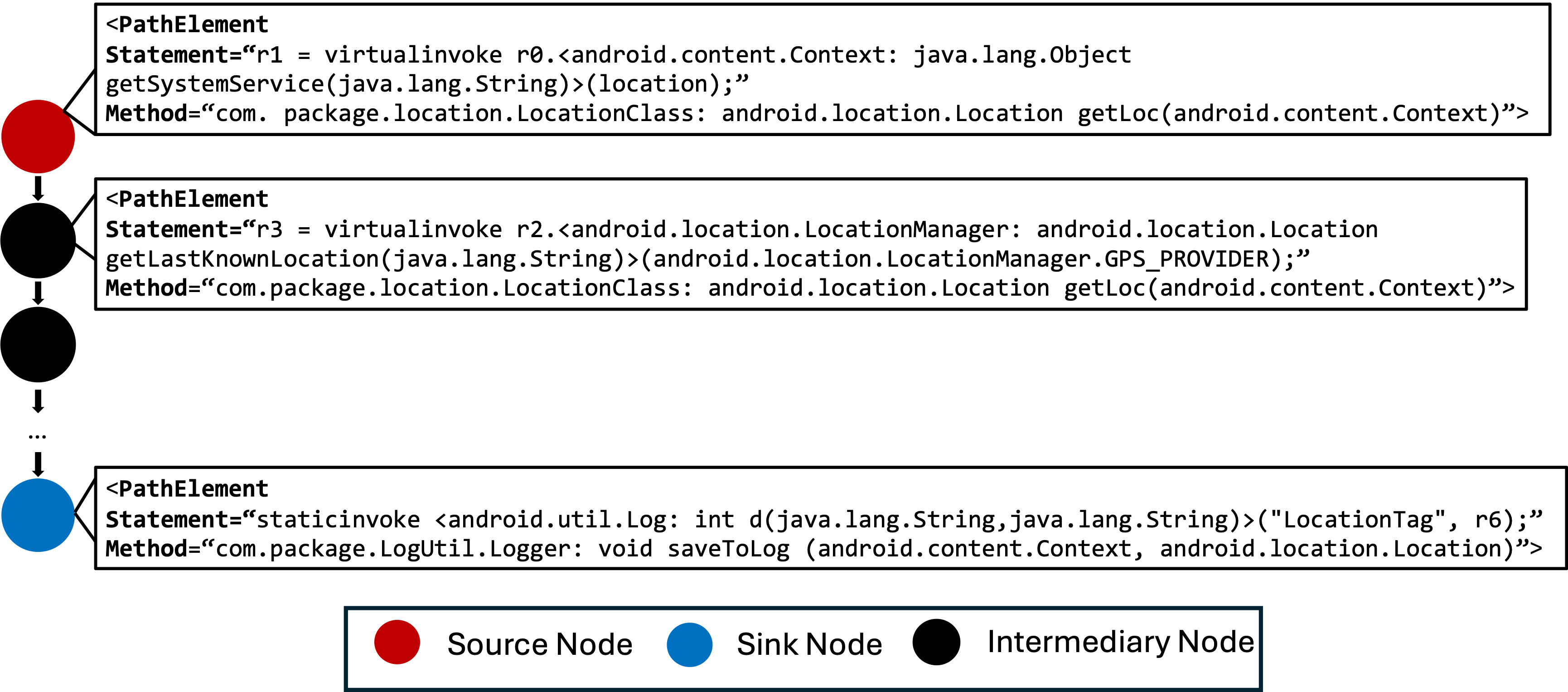}
    \caption{An example of a taint path describing the flow of location information from the source method \texttt{getLoc} being saved to a log file in the sink method \texttt{saveToLog}.}
    \label{fig:taint-path}
\end{figure}

FlowDroid 
identifies data-flow paths between source methods (such as \texttt{getLastKnownLocation}) and sink methods (such as \texttt{sendTextMessage}) that are pre-defined in a list (\texttt{SourcesAndSinks} file\footnote{\url{https://tinyurl.com/SourcesAndSinks}}). The list contains several location-based Android APIs and other commonly used APIs, such as for saving files or sending data over HTTP. FlowDroid models the Android lifecycle to identify potential entry points (e.g., \texttt{OnCreate()} method), builds a control-flow graph of the app, and tracks how data moves across methods using interprocedural analysis. When data originating from a listed source method 
reaches a listed sink method, FlowDroid traces the sequence of statements and method invocations involved in the data flow, and outputs all source-to-sink paths in an XML file. 

Figure \ref{fig:taint-path} shows part of an extracted taint path. Each path consists of several nodes, starting from the source node and ending with the sink node, where each node in the taint path describes in which statement (i.e., the \texttt{Statement} field in each node in Figure \ref{fig:taint-path}) and in which method and class (the \texttt{Method} field in the nodes) the data flows. Note that the statements in the \texttt{Statement} attribute are in Jimple format, an intermediary representation of Java source code \cite{vallee2010soot}, which are not human-readable.


While simple, a primary challenge with using taint analysis for extracting taint paths is that it can only extract paths between the \emph{listed} source and sink methods and cannot do so for unlisted methods. While we can easily add source methods based on Android documentation as new APIs are released, adding sink methods becomes very challenging. Consider an example where an app uses the method \texttt{getConnectionInfo} to extract WiFi information, such as the MAC address of an access point or an IP address. This method is not listed in the original file, but can be added. While the app may not use this information through one of the listed sink APIs, such as saving it to a file or sending it over HTTP, it may use the information in its own method that is not prelisted.
Since we cannot know apriori which method in the app uses sensitive personal information, it becomes non-trivial to determine the taint path from a known source method to an unknown sink method. As FlowDroid constructs a data-flow graph, in theory, it is possible to find all sink methods from a new source method, but this is practically infeasible due to computational complexity. 

We address the limitation of traditional taint analysis by developing a novel approach using static analysis and heuristics to extract taint paths between known sources and unknown sink methods.
We first construct a call graph for a given source API (e.g., \texttt{getConnectionInfo}) up to a fixed depth and treat all methods (at all depths) in the call graph as potential sink candidates. We then list all these potential sinks and run taint analysis to extract the taint paths if they exist. This selective strategy restricts the search space and reduces the computational cost of exploring all possible source-sink method combinations, while allowing us to recover meaningful taint paths that describe the use of sensitive personal information. 

A previous study found that in 80\% of cases, personal information is used within the first three methods of a call graph starting from the source method~\cite{jain2022pact}. To determine the optimal call graph depth, we performed a breadth-first search across 50 Android apps to examine the distribution of sink nodes at different levels. We found that most sinks appear within four to five method calls from the source, with very few occurring beyond the sixth level. Therefore, to balance the trade-off between accuracy and efficiency, we use a depth of six to construct the call graph. Note that the maximum depth of a call graph we found was 10.

Since FlowDroid's taint analysis already works well for listed location-based APIs, we analyzed the efficacy of our modified approach with APIs from other categories, such as camera, storage, WiFi, sensors, and account.  This totaled 80 unlisted (in FlowDroid) non-location API sources.


We identified taint paths from 500 apps by dedicating 16 GB of memory per app when running its taint analysis. This memory limit was empirically chosen based on prior taint analysis runs for location APIs, which only needed 8GB of memory. We doubled the memory limit to improve accuracy. 
After analyzing 500 apps (that took  $\sim$60 hours on one computer with an AMD Ryzen 9 CPU and 124 GB of RAM), we found valid taint paths associated with only 10 (out of 80) APIs from the categories of WiFi, account, and network. By a valid taint path, we mean paths that contain at least three nodes in the flow, since this is the threshold number of methods for processing personal information in a call graph and describing a privacy behavior~\cite{jain2022pact}. 

These results indicate that our modified approach, although not perfect, is highly effective for WiFi, network, and account-related APIs -- which are used in apps to access sensitive personal information such as IP address, MAC address, and other account-related information, that can be used to fingerprint devices and track users. However, our modified approach is not effective for camera, sensors, and storage APIs. This discrepancy indicates that the flow of personal information differs based on the API category, where some 
categories may pass the information through components or asynchronous callbacks that are not part of the call graph and therefore cannot be obtained via our modified approach. Despite this drawback, our approach complements taint analysis in FlowDroid by identifying sinks unique to each app and discovering new sinks that are not known and listed apriori -- expanding the capabilities of traditional taint analysis approaches that rely on known sources and sinks.

\subsection{Extractor: Extracting Source Code}
\label{subsec:extractor}

This component extracts source code of the classes, methods, and statements in the taint paths, because 
taint paths alone do not provide the complete context and are insufficient for an LLM to generate privacy captions. \textit{Extractor} accomplishes this task using static analysis and LLMs.


There are two key challenges for extracting source code from taint paths: first, extraction needs to happen from compiled APK files instead of open-source projects, and we cannot easily search the classes and methods involved in the taint path. Second, the source code of the involved methods provides only a coarse-grained context for an LLM, which may not generate concise, complete, and accurate privacy captions, since each taint path contains several methods and classes, thereby creating a very large source code context. For a fine-grained context, we need to provide the statements through which personal information flows. These statements are depicted in the taint paths in Jimple format, which is not human-readable; therefore, we must map them to the corresponding statements in the source code. 

We address the first challenge by using a reverse engineering tool to decompile the APK files and extract the source code. Specifically, we use Androguard \cite{androguard} since it has been used by previous works \cite{jain2022pact, jain2023towards, martin2018new}. 
We process the string in the Method attribute of each taint path node, which is in the format of \texttt{<class\_name>: <return\_type><method\_name> <argument\_types>}, to identify the involved methods and classes. Using this signature, we extract the source code of the methods from the APK file using Androguard. We also concatenate the 
methods that belong to the same class, and write the merged source code as a `.java` file. Therefore, we transform each taint path into a collection of class files, where each file contains the source code of methods from the taint path. A notable advantage of using a reverse engineering tool is that it can handle both open and closed-source apps and even extract the source code of third-party libraries that are not accessible. 

To address the second challenge, we tested multiple approaches to map the Jimple statements to the extracted source code. Specifically, we tried similarity search with heuristics-based matching; however, since the extracted source code is obfuscated, it compounds the complexity of matching Jimple statements. We then leveraged LLMs to map the statements and found good results. We experimented with several models: GPT-4 (gpt-4-0613) \cite{achiam2023gpt}, Claude 4 Opus (claude-opus-4-20250514) \cite{claudeopus4}, LLaMA 3 70B (llama-3-70b-8192) \cite{dubey2024llama}, and DeepSeek (deepseek-r1-distill-llama-70b)~\cite{liu2024deepseek} -- a mix of leading proprietary and open-source models. Apart from LlaMa 3, all models performed well at this task. We finally selected DeepSeek, as we could run the model locally, minimizing our environmental footprint and API costs. We designed the prompt based on In-Context Learning (ICL) \cite{brown2020language} as we saw the model follows the instructions better and learns from previous examples shared in the prompt. We show the final prompt we used to map the statements in Appendix \ref{appendix:map-jimple-prompt}.

Extractor transforms the extracted taint paths into the source code of methods and classes in the paths, along with annotations of the statements through which personal information flows. This output source code provides both \textit{coarse-grained} and \textit{fine-grained} context for an LLM to comprehend privacy behaviors and generate privacy captions.

\subsection{Generator: Generating Privacy Captions}
\label{subsec:generator}

After identifying the taint paths and then extracting the source code, the last step in PCapGen is to generate privacy captions that describe the privacy behaviors implemented in the source code context. We use LLMs for this task because they can handle large source code contexts and have demonstrated strong performance in similar commit message generation tasks \cite{li2024only, eliseeva2023commit}. 

We iteratively refined and developed our prompt by employing various prompt engineering techniques, such as prompt chaining and few-shot learning.
For this,
we downloaded 10 random apps from the AndrooZoo dataset \cite{Allix:2016:ACM:2901739.2903508} and created 20 source code samples using 
the steps described in Sections \ref{subsec:identifier} and \ref{subsec:extractor}. We then manually created privacy captions for these samples and used them for comparison. We compared the caption generation performance across four models - GPT-4 (gpt-4-0613) \cite{achiam2023gpt}, Claude 4 Opus (claude-opus-4-20250514) \cite{claudeopus4}, LLaMA 3 70B (llama-3-70b-8192) \cite{dubey2024llama}, and DeepSeek (deepseek-r1-distill-llama-70b)~\cite{liu2024deepseek}. 
In our evaluation, we found that LlaMa 3 often generated inaccurate captions (created captions for the example in the prompt, but not for the actual code sample), so we skipped LlaMa 3 for this task. 

Creating privacy captions is a complex task where the model needs to: i) comprehend the source code in multiple class files and understand how the data flows between statements, methods, and classes; ii) understand the privacy behaviors implemented in these multiple files; and iii) summarize them in a simple caption. We initially began our process of iterative prompt refinement with the \textit{prompt chaining} technique \cite{kwakclassify2024}, where we broke down the privacy caption generation into these three sub-tasks. This made intuitive sense because 
having sub-tasks reduces the probability of error and allows for traceability 
as we can pinpoint which step caused the erroneous output.
In the first task, we provide the model with the source code and taint analysis statements as context and instruct it to identify relevant statements (beyond what taint analysis has identified) that use personal information. 
In the second task, we provide the model with the source code context and the identified relevant statements, and instruct it to create privacy labels, specifically PAcT labels \cite{jain2022pact}, to explain how personal information is used in the source code context and why. In the last step, we prompt the model to generate a privacy caption using the source code context, identified relevant statements, and privacy labels. For all sub-tasks, the prompts followed in-context learning (ICL) \cite{brown2020language} approach and included examples to help the model follow the instructions more effectively.



While this prompt chaining approach seemed promising, we encountered several issues during evaluation. First, while GPT, Claude, and DeepSeek successfully identified relevant statements, LLaMa often hallucinated statements that did not exist in the source code. Second, the generated privacy labels were often incorrect, particularly when multiple labels were involved. Third, the final privacy captions generated were generic and essentially used labels as templates.
This prevented the model from creating captions that describe the nuances of privacy behavior. Lastly, the multi-step process also introduced error propagation, where errors in early sub-tasks compounded in subsequent outputs. 

Based on these observations, we pivoted to a simpler approach, where
we designed a single prompt that provides the model with the complete context needed to generate accurate privacy captions. In this version, following in-context learning (ICL) \cite{brown2020language}, we provided the model with the source code of the sample, the taint analysis statements, and an example tuple of source code and its privacy caption, and asked it to generate a privacy caption for the code sample. We obtained positive results for small and medium-sized samples, but the privacy captions for large samples did not capture all privacy behaviors implemented in the source code. Therefore, we selected different examples for each size and provided them accordingly. This dynamic (example changing based on the size of the code sample) in-context learning prompt yielded good results during evaluation, which is shown in Appendix \ref{appendix:pcapgen-prompt}. Since GPT4, Claude, and DeepSeek all generated good captions, we consider all three models for our experimentation and consider them as 
configurations of PCapGen with labels: 
PCapGen\_GPT, PCapGen\_Claude, and PCapGen\_DeepSeek respectively.

\label{sec:approach}

\section{Baseline Captions, Survey, and Dataset}

To evaluate the efficacy of PCapGen in generating privacy captions, we need a dataset of Android source code samples and their privacy captions. However, no such dataset exists. 
A significant challenge in creating one is to verify the correctness and quality of the privacy captions. In this section, we present our methodology for creating such a dataset. 
We begin by detailing the process of extracting Android code samples using the static and taint analysis approaches as discussed in Sections \ref{subsec:identifier} and \ref{subsec:extractor}. Next, we outline our baseline approach for generating privacy captions, which we refer to as \textit{baseline privacy captions}. To assess the quality and correctness of the captions, we conduct a survey with professional Java developers. We describe the survey design, the pilot study findings, and subsequent modifications to the survey. Lastly, we discuss the characteristics of the final dataset of 442 unique Android source code samples and their developer-verified baseline privacy captions. 

\subsection{Extracting Android Source Code Samples}
\label{subsec:create-src-code-samples}

To extract Android source code samples, we downloaded 11,173 APK files from the Androzoo collection \cite{Allix:2016:ACM:2901739.2903508}, as it is 
the largest archive of Android apps from various app marketplaces and is commonly used in prior research~\cite{zhao2025apppoet, nayan2024sok}.
We then used our taint analysis approach described in Section \ref{subsec:identifier} to extract over 15,000 code samples. To ensure uniqueness among samples, we employed the MinHash algorithm \cite{broder1998min} with a Jaccard similarity threshold of 0.5. This threshold suggests that samples with at least 50\% similarity are considered duplicates. We selected this threshold after conducting an empirical analysis that showed it removed near-duplicate code samples while preserving diversity among them. A threshold of 0.7 or 0.8 retained too many similar samples, whereas a threshold of 0.3 drastically reduced the sample count and variations in the behaviors. Using this threshold, we curated 589 unique code samples, comprising 383 samples from location-based APIs and 206 samples from non-location-based APIs.  

\subsection{Baseline Approach to Create Privacy Captions}
\label{subsec:baseline-approach}


Ideally, we would ask developers to create privacy captions for the curated source code samples and use these as a baseline to compare with the PCapGen-generated captions.
However, asking developers to write the baseline privacy captions after reading code samples spanning hundreds of lines is time-consuming, complex, expensive, and (sometimes) error-prone, making such an approach infeasible for creating a large dataset. In comparison, it is a simpler task to verify a caption generated for a source code sample. Therefore, we use LLMs as a proxy for developers to generate these baseline captions, and then we validate the generated captions using developers. 
We design the prompt by mimicking how a developer would create these privacy captions and what information can be shared with them.
The prompt provides detailed instructions and only the source code sample as the context, without any taint analysis statements. 
We show the baseline prompt in Appendix \ref{appendix:baseline-prompt}.

We conducted a comparative evaluation to select the most suitable LLM for generating baseline captions. We compared GPT-4 (gpt-4-0613) \cite{achiam2023gpt}, Claude 3 Opus (claude-3-opus-20240229) \cite{claudeopus3}, LLaMA 3 (llama-3-70b-8192) \cite{dubey2024llama}, and DeepSeek (deepseek-v2)~\cite{liu2024deepseek}. 
We generated captions for 20 randomly selected source code samples using each model and manually evaluated the captions on their correctness and quality. We found GPT-4 produced more accurate captions that described the privacy behaviors in detail. Based on these results, we selected GPT-4 to generate baseline captions for all 589 unique code samples in our baseline dataset. We used a temperature of 0.7 to generate the baseline captions because it has been used in other works as well \cite{zhengjudging2023}.

\subsection{Survey Design}
\label{subsec:survey-design}

\begin{figure*}[t]
  \centering

  \begin{subfigure}[t]{0.48\textwidth}
    \centering
    \includegraphics[width=\linewidth]{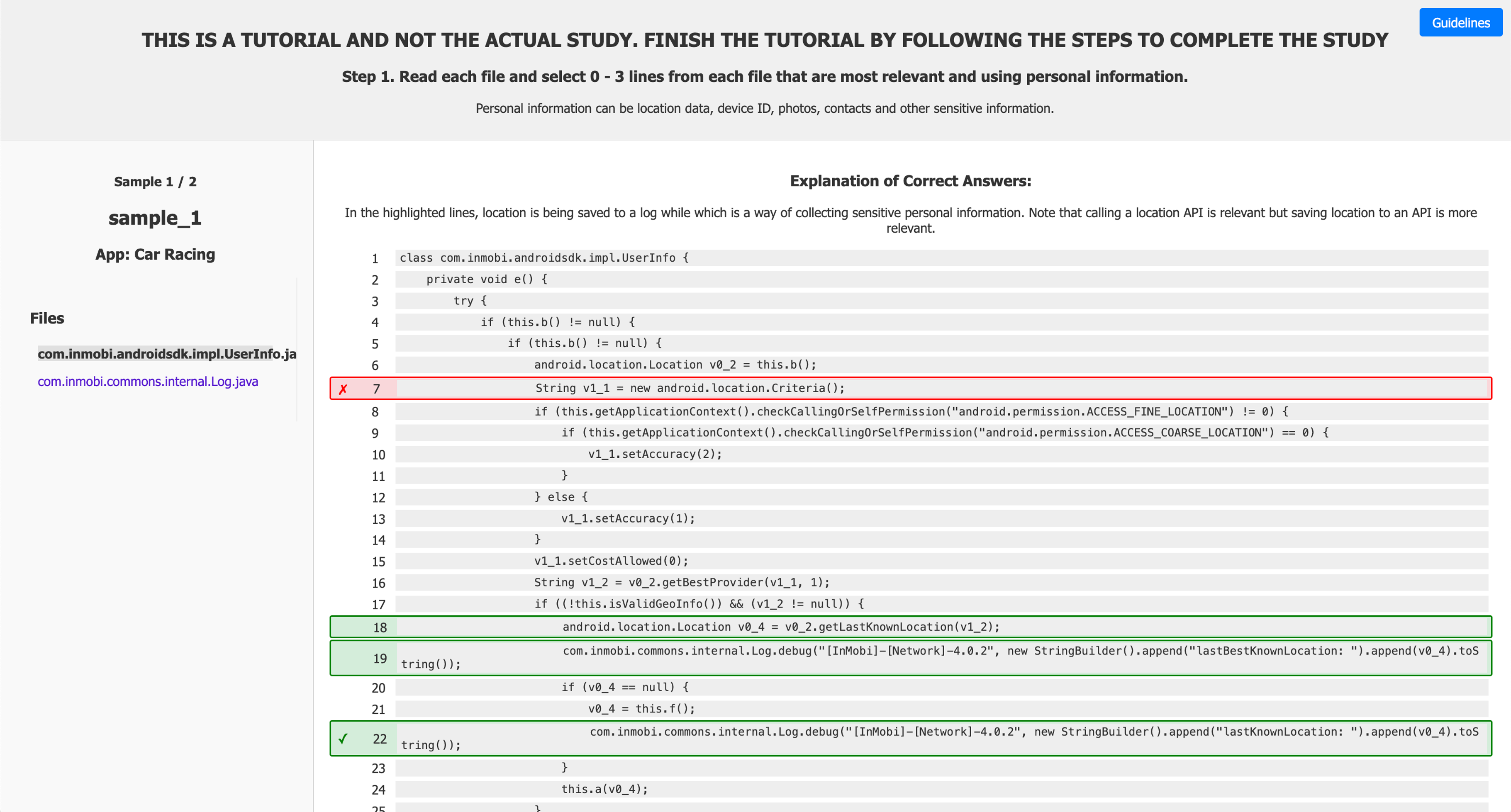}
    \caption{Step 0. Onboarding annotators and showing them correct responses with an explanation}
    \label{fig:step-0}
  \end{subfigure}
  \hfill
  \begin{subfigure}[t]{0.48\textwidth}
    \centering
    \includegraphics[width=\linewidth]{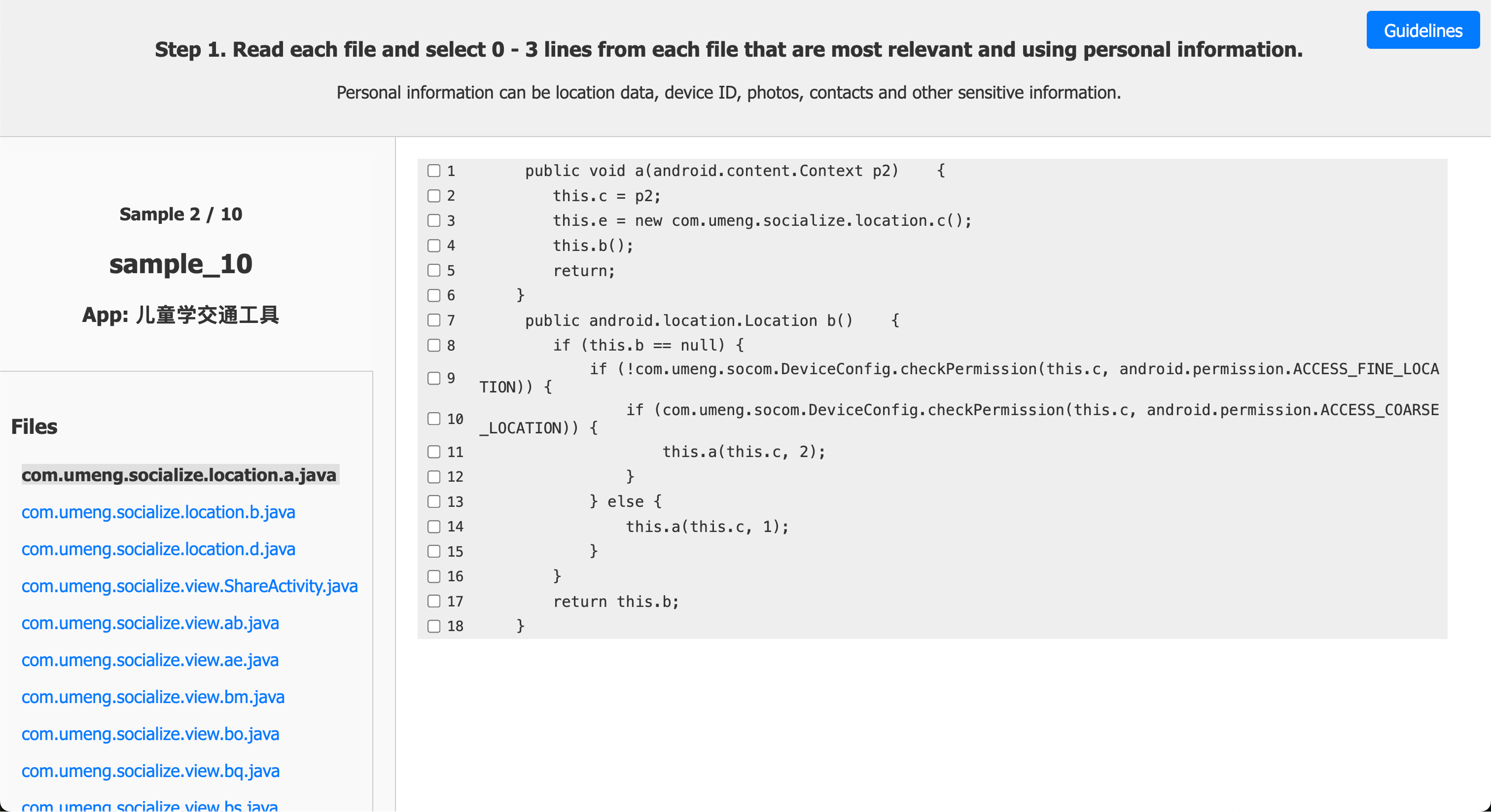}
    \caption{Step 1. Asking annotators to comprehend the entire source code and select relevant statements}
    \label{fig:step-1}
    
  \end{subfigure}

  \vspace{0.75em}

  \begin{subfigure}[t]{0.48\textwidth}
    \centering
    \includegraphics[width=\linewidth]{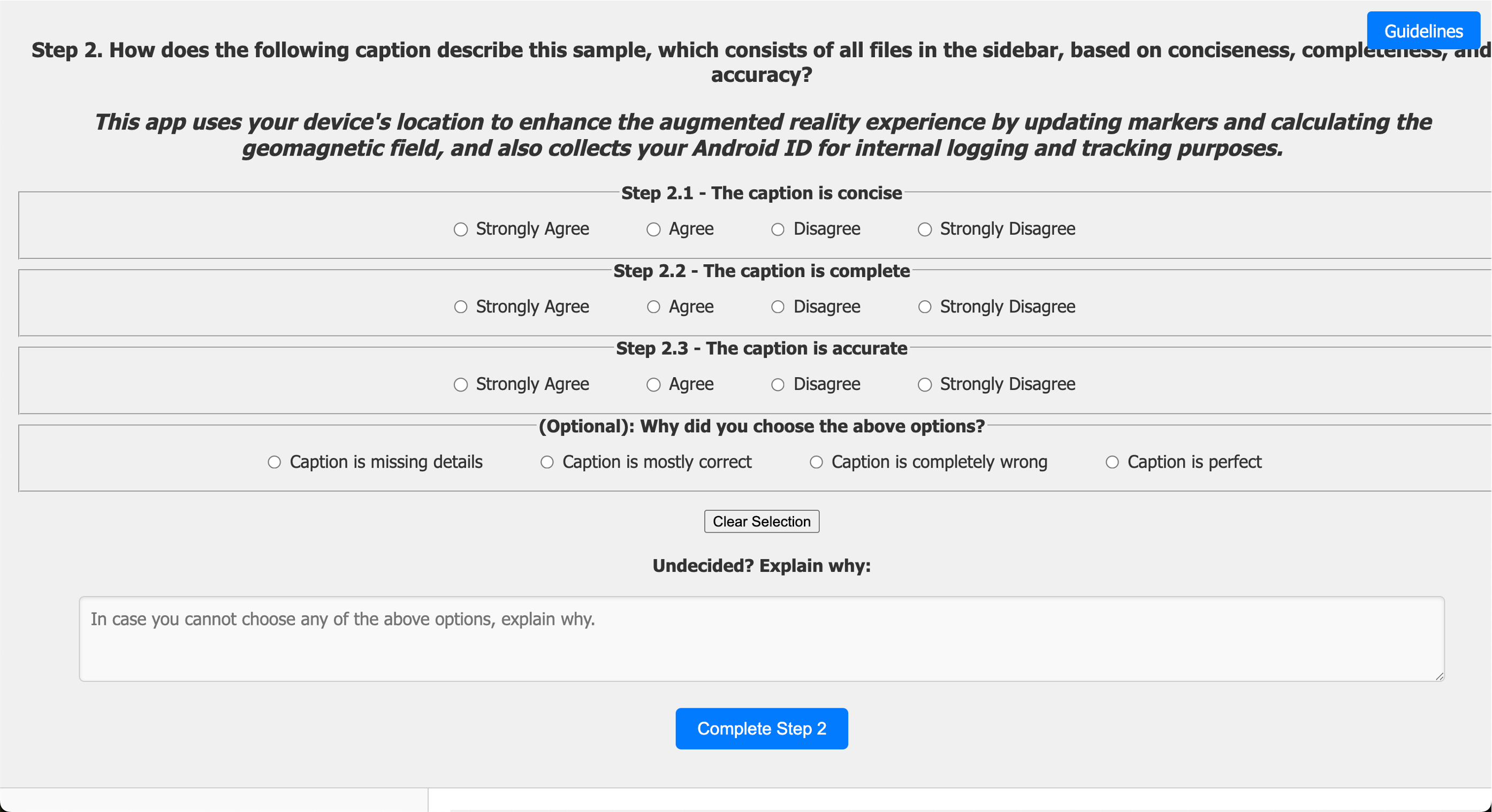}
    \caption{Step 2. Asking annotators to rate baseline caption}
    \label{fig:step-2}
  \end{subfigure}
  \hfill
  \begin{subfigure}[t]{0.48\textwidth}
    \centering
    \includegraphics[width=\linewidth]{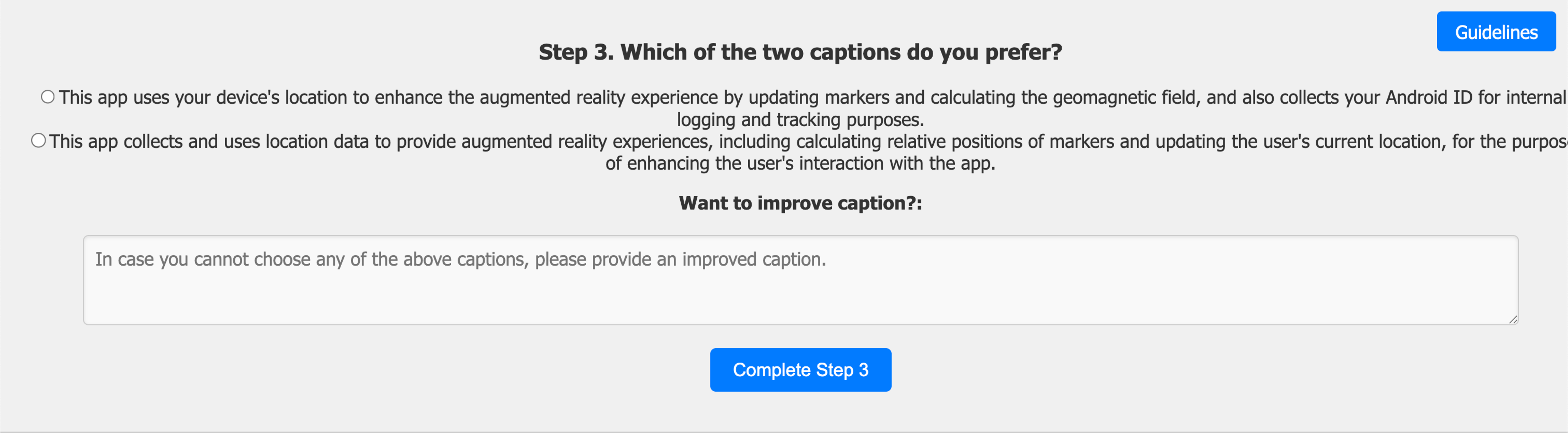}
    \caption{Step 3. Asking annotators to choose the better caption}
    \label{fig:step-3}
  \end{subfigure}

  \caption{User Interface of Final Annotation Tool.}
  \label{fig:annotation-tool-ui}
\end{figure*}

After generating baseline LLM captions, 
we design a survey to ask professional developers to verify their correctness and quality. Specifically, we ask Java developers to rate the privacy caption on its quality and improve it. We base this design on a similar work evaluating code captions for Java methods \cite{su2024distilled}. The survey has three parts: i) comprehending the source code sample (as shown in Figure \ref{fig:step-1}), ii) rating the LLM generated privacy caption (Figure \ref{fig:step-2}), and iii) improving the caption, if there are any issues with it (Figure \ref{fig:step-3}). Before beginning the survey, we provide developers with an interactive demo of these three steps. In this demo, developers are provided with instructions and asked to perform each of the three tasks for a code sample. After they answer each step, they are provided with the correct answers and detailed explanations to help them understand the task better. Figure \ref{fig:step-0} shows the demo for Step 0. A guideline box is always accessible to the developers for guidance on any step. 

In Step 1 of the survey (Figure \ref{fig:step-1}), developers are asked to read the source code in each file and select statements that describe the use of personal information. Recall that each sample contains multiple class files, and each class file can have multiple methods. 
In this step, developers are required to actively read and comprehend these source code samples, as each sample contains several interconnected methods that require focused attention. The developer-selected statements can be used to evaluate whether the developers actually comprehended the source code or selected statements randomly. For example, in case the developer chooses meaningless statements, such as return statements or curly braces, we can discard those annotations. In Step 2 (Figure \ref{fig:step-2}), developers are asked to rate the accuracy (we later added conciseness and completeness criteria based on the pilot study -- Section~\ref{subsec:pilot}) of the baseline caption from four options -- ``Strongly Disagree", ``Disagree'', ``Agree'', and ``Strongly Agree". We did not have a neutral option to nudge developers to provide a rating \cite{su2024distilled}, however, we provided a text box where developers could explain their reasons if they could not make any selection. We also included an optional question to  allow developers to explain their ratings. 
If developers rate the caption as ``Disagree'' or ``Strongly Disagree,'' then in Step 3 (see Figure \ref{fig:step-3}), they are required to improve the initial privacy caption. We developed our own annotation tool for this survey. 
To avoid any bias, each code sample gets annotated by two developers.

\subsection{Pilot Study}
\label{subsec:pilot}

We first conducted a pilot study with two software developers on Prolific to evaluate the baseline caption rating task. We recruited participants who had at least 6 years of professional experience in programming with Java, were fluent in English, and had an approval rating of more than 95\% on the platform.  Based on the pilot study findings, we modified our research method as follows. First, we found that participants selected several statements in each file that included irrelevant lines, such as curly braces or return statements. To restrain this practice of selecting non-useful lines, we limited the selection to at most three statements per file. Second, we found that accuracy alone did not capture all the dimensions of a caption's quality; therefore, we revised the survey and asked the developers to rate the caption on three criteria: conciseness, completeness, and accuracy. These three criteria have been used previously in a similar task \cite{su2024distilled}. Third, we also realized that participants would skip rating the caption with ``Disagree" or ``Strongly Disagree" in Step 2, as it would require them to write a new privacy caption, which is a more intensive task compared to choosing a caption. Thus, we generated a second version of the privacy caption for each sample and asked the developers to choose between the two captions. We generated this second caption using the same baseline prompt and temperature setting, and used the same model as before. Generating two versions of the output using the same temperature and model configurations for comparison has been used in prior work~\cite{zhengjudging2023}.

We still provided a textbox to update the privacy caption in case none of the options were relevant. Lastly, since each sample would have two versions of the baseline privacy caption and we have two developers rating each code sample, we modify Steps 2 and 3 as follows. In Step 2, one developer would rate the first version of the caption, while the second developer would rate the second version; this would give us individual ratings for both captions. In Step 3, both developers are shown both the versions and are asked to choose the better one. The order of captions would be shuffled to avoid any positional bias. In this modification, both developers get to evaluate both the caption versions before providing their preference.








\subsection{Participants and Baseline Caption Dataset}
\label{subsec:participants}

After the pilot study, we used Prolific's Domain Experts, who are verified professionals with expertise \cite{Kopić_Engineer}, to hire software developers with experience in Java. We also required them to have an approval rating of 95-100\% on the platform and be fluent in English. To further improve the quality of participants, we established a screening process where those meeting our base criteria were invited to annotate 10 samples. We manually reviewed the annotations of these 10 samples to assess whether the participants selected relevant lines demonstrating comprehension of source code and correctly rated the privacy captions showing understanding of the implemented privacy behaviors. If the annotations were correct, we invited them to the actual study. If the annotations were incorrect, such as the wrong selection of statements or exceptionally quick completion, we discarded the annotations and did not invite the participants to the study. Out of 50 eligible participants who met the base criteria, 30 participated in the screening study, and ultimately, 16 were invited again to annotate the 589 dataset samples. In each annotation session, participants were asked to annotate 10 samples, which took a median time of 40 minutes. We paid them \$12  for each session, which translates to approximately \$18/hour, which is considered a good rate as per Prolific's guidelines.

After the final study, each of the 589 samples contained two versions of the privacy caption, ratings for each version by one participant, and preferences between the two versions by two participants. We filtered the baseline captions and samples using the following criteria: 
We first ensured that each caption version was rated at least as `Agree' on all three quality dimensions (i.e., conciseness, completeness, and accuracy). Captions that did not satisfy this condition were discarded. In cases where both caption versions for a sample did not satisfy the condition, the entire sample was discarded. Next, if both caption versions were rated with at least `Agree' on all dimensions, then the caption that both participants chose as their preferred one was selected as the baseline privacy caption for the sample. In cases where there was no clear preference between the two versions, we used both as the baseline privacy caption for the sample, since participants may have subjective opinions about quality. After applying this filtering process, we removed 147 samples, resulting in a final dataset of 442 unique source code samples and their baseline captions, of which 54 had 2 baseline captions.

\label{sec:dataset}

\section{Evaluation Experiments}

In this Section, we describe our approach to evaluate PCapGen captions as compared to the baseline captions. We ask human privacy experts and LLM-as-judge models (referred to as judges) to compare captions both quantitatively and qualitatively. Quantitatively, both experts and judges evaluate captions on key quality criteria, and qualitatively, they compare the two captions and indicate their preference. We propose using LLMs-as-judge because these models can provide comparable evaluations to humans and allow us to scale to many more apps, compared to just a few samples with humans. Also, LLMs are turning into a cost and time-effective evaluation tools that are increasingly being used in qualitative analysis \cite{robertsartificial2024, demoraislecaapplications2025, banoai2023}. In our experiments, we use three judges, specifically -  LlaMa 3.3 (llama-3.3-70b-instruct) \cite{Llama_3.3}, Mistral \cite{mistral_ai} (mistral-medium-3), and Gemini 2.5 Pro \cite{comanici2025gemini} (gemini-2.5-pro), as a mix between open-source and proprietary models. We specifically avoided GPT, Claude, and DeepSeek models because LLMs are known to have self-preference bias where they promote specific styles or policies intrinsic to the LLM~\cite{wataoka2024self}.
We evaluate PCapGen using the following research questions outlined in Section \ref{sec:introduction}.



\subsection{RQ1: Comparing PCapGen Configurations}

The goal of RQ 1 is to compare all configurations of PCapGen, i.e., PCapGen\_GPT, PCapGen\_Claude, and PCapGen\_DeepSeek, and identify the configuration that generates the best captions as per the quality criteria. To evaluate every caption, we ask the three LLM judges to rate the privacy caption on three quality criteria -- accuracy, conciseness, and completeness -- on a scale of 1 - 5~\footnote{Unlike in the baseline caption rating task (Figure~\ref{fig:step-2}), we include the ``Neutral'' option here}, where 1 is ``Strongly Disagree'' and 5 is ``Strongly Agree'' \cite{gao2023human, jones2024multi}. 
The primary reason for evaluating LLM-judges is to validate whether the caption quality evaluation can be automated and scaled. 
We design our prompt based on guidelines provided in prior work to use LLMs-as-judge for qualitative evaluation \cite{gusurvey2025}. We show the prompt in Appendix \ref{appendix: second-eval-prompt}.

\subsection{RQ2: Comparing Baseline and PCapGen}
\label{sec:semantic-similarity-approach}
The RQ 2 goal is to evaluate the semantic differences between the two captions - baseline and PCapGen. Semantic similarity/difference between them provides an initial analysis of whether PCapGen generates captions that are as good as or better than the developer-curated baseline captions. 
High semantic similarity may indicate that the two captions are very similar and discuss the same aspects. 
Whereas low similarity may indicate differences between the two, and we need to look at the quality criteria to determine if PCapGen captions are an improvement over the baseline captions. We evaluate semantic similarity using privacy experts (researchers from our lab with experience working on privacy), LLMs-as-judge, and automated metrics. With human experts, we randomly select 90 samples (30 samples for each LLM model, i.e., Claude, DeepSeek, and GPT) and ask two privacy experts to rate the similarity on a scale of 1 - 5, where 1 is completely dissimilar and 5 is identical. We ask the three judges (i.e., LLMs-as-judge models) to do the same for all samples in the dataset. We provide the experts and the judges with a rubric to guide them in rating similarity, so that the rating values are interpreted consistently. We  use automated similarity metrics, e.g., BERTScore \cite{zhang2019bertscore}, BARTScore \cite{yuan2021bartscore}, and ROUGE-L \cite{linrouge2004}, to quantitatively assess the similarity between PCapGen and baseline captions. BERTScore and BARTScore have been used to compare the semantic similarity of code captions~\cite{haquesemantic2022}, whereas ROUGE-L provides an n-gram measure of overlap between the two captions. In instances where a code sample has two baseline captions, we average their answer scores.

\subsection{RQ3: Comparing Captions with Judges}

Here, we compare privacy captions from the best PCapGen configuration and the baseline captions using LLM-as-judge models. The best configuration of PCapGen is identified from RQ 1 earlier, and similar to the task in RQ 1, we ask the judges to rate both PCapGen and baseline captions separately on three key quality criteria - conciseness, completeness, and accuracy. The judges are prompted to rate each caption on a scale of 1-5 for each criterion using the prompt listed in Appendix \ref{appendix:first-eval-prompt}. While these metrics are helpful, they alone are not sufficient to provide decisive comparative results, as indicated by previous works \cite{Stapleton2020Humanstudy, roy2021reassessing}. Hence, we also ask the judges to choose the better caption between the two options. We shuffle the caption order in the prompt to avoid any positional bias~\cite{lillmsasjudges2024}.

\subsection{RQ4: Comparing Captions with Experts}
\label{sec:expert_survey_method}
Here, we ask the two privacy experts from earlier (Section~\ref{sec:semantic-similarity-approach}) to evaluate the captions from the best PCapGen configuration and the baseline captions on the same three criteria 
on a scale of 1 - 5. We designed this task very similar to our developer annotation survey in Section~\ref{subsec:survey-design}. Privacy experts are first asked to comprehend the source code sample, rate the privacy caption as per each criterion on a scale of 1-5~\footnote{Unlike in the baseline caption rating task with developers, for privacy experts, we include the `Neutral' rating option in Figure~\ref{fig:step-2}. This is to ensure parity between the LLM-judge and human privacy expert settings.}, 
and then choose the better caption between the baseline and PCapGen. When evaluating with judges, we provide the source code and the taint analysis statements as the context. To ensure parity of information shared across the judges and human privacy expert settings, in the first step (Figure~\ref{fig:step-1}), experts are also shown highlighted taint analysis statements in the source code. The best configuration of PCapGen is identified from RQ 1. We select 120 random code samples along with their corresponding best PCapGen and baseline captions for this evaluation task. Across 60 samples, experts are asked to rate the PCapGen captions based on the quality criteria, and in the other 60 samples, they rate the baseline captions. Note that across all these 120 samples, experts give their preferred caption between the PCapGen and baseline versions (Figure~\ref{fig:step-3}).

\label{sec:prompt-engg-exp}

\section{Results}

\begin{table*}[h]
\centering
\caption{Average ratings for conciseness, completeness, and accuracy by LlaMa, Gemini, and Mistral judges}
\begin{tabular}{|l|r|r|r|r|r|r|r|r|r|}
\hline
\textbf{Configuration} & \multicolumn{3}{c|}{\textbf{Conciseness}} & \multicolumn{3}{c|}{\textbf{Completeness}} & \multicolumn{3}{c|}{\textbf{Accuracy}} \\ \hline
 & \textbf{LlaMa} & \textbf{Gemini} & \textbf{Mistral} & \textbf{LlaMa} & \textbf{Gemini} & \textbf{Mistral} & \textbf{LlaMa} & \textbf{Gemini} & \textbf{Mistral} \\ \hline
Baseline & 4.037 & 3.893 & 3.900 & 4.009 & 2.904 & 3.954 & 4.529 & 3.477 & 4.253 \\ \hline
PCapGen\_Claude & \textbf{4.344} & \textbf{4.535} & 4.025 & \textbf{4.380} & \textbf{3.803} & \textbf{4.095} & \textbf{4.913} & \textbf{4.522} & \textbf{4.701} \\ \hline
PCapGen\_DeepSeek & 4.214 & 4.378 & \textbf{4.073} & 4.079 & 3.579 & 4.039 & 4.735 & 4.220 & 4.459 \\ \hline
PCapGen\_GPT & 4.138 & 4.197 & 4.012 & 4.080 & 3.357 & 4.022 & 4.802 & 4.185 & 4.528 \\ \hline
\end{tabular}
\label{tab:llm-rating}
\end{table*}

In this section, we discuss our evaluation results for our experiments described in the previous section. 

\subsection{RQ1: Comparing PCapGen Configurations}
\label{subsec:config_comparison}

We report the ratings for each configuration across each criterion and by each judge in Table~\ref{tab:llm-rating}. Note that the values in each cell are averaged across all 442 source code samples in the baseline dataset. Across all three criteria, values close to 5 (`Strongly Agree') are better. From the table, we can see that PCapGen\_Claude consistently performs better than other configurations, according to all judges (Llama, Gemini, and Mistral). Only for the conciseness criterion, Mistral prefers PCapGen\_DeepSeek, but PCapGen\_Claude is not far behind. 

To validate if the differences in the LLM judge ratings across all configurations are statistically significantly different, we run paired t-tests between all three configuration pairs. The results are shown in Table~\ref{tab:pcapgen_comparisons}. We perform pairwise comparison of the configurations by averaging a sample's ratings received over all quality criteria (i.e.,  accuracy, conciseness, and completeness) and for all three judges. The null hypothesis is that the mean of the differences between the paired observations is zero. We adjust the p-values based on the Holm-Bonferroni method to counteract the problem of multiple comparisons. In Table~\ref{tab:pcapgen_comparisons}, the \textit{comparison} column indicates which two configurations are compared, \textit{mean\_diff} is the difference in the average rating across the two configurations, \textit{$p\_holm$} is the adjusted p-value, and \textit{$Cohens\_d_z$} indicates the effect size or the significance of the difference between the pairs (a magnitude of 0.2-0.5 indicates small effect, 0.5-0.8 indicates medium effect, and $>=0.8$ indicates large effect; the positive or negative sign indicates which mean is greater).

As Table \ref{tab:pcapgen_comparisons} indicates, PCapGen\_Claude has higher average ratings across all criteria and judges compared to PCapGen\_GPT and PCapGen\_DeepSeek, with $|mean\_diff|>0.19$. The $p_{holm} = 0$ and $|Cohens\_d_z|>0.5$ indicate statistical significance and medium effect size. 

When comparing PCapGen\_DeepSeek and PCapGen\_GPT, we do not find any statistically significant difference between the two configurations, though Table~\ref{tab:llm-rating} shows PCapGen\_DeepSeek scoring slightly higher than PCapGen\_GPT in most criteria. 

In conclusion, these results show that PCapGen\_Claude performs the best in all criteria, and we choose this as the best PCapGen configuration for generating privacy captions. 

\begin{table*}[h]
\centering
\caption{Pairwise comparison of PCapGen configurations averaged over all quality criteria}
\begin{tabular}{|l|r|r|r|}
\hline
\textbf{Comparison} & \textbf{mean\_diff (b--a)} & \textbf{$p_{holm}$} & \textbf{$Cohen's \_d_z$} \\ \hline
PCapGen\_Claude v/s PCapGen\_DeepSeek & -0.191781 & 0 & -0.54 \\ \hline
PCapGen\_GPT v/s PCapGen\_Claude      &  0.197869 & 0 &  0.559 \\ \hline
PCapGen\_GPT v/s PCapGen\_DeepSeek    &  0.006088 & 1 &  0.016 \\ \hline
\end{tabular}
\label{tab:pcapgen_comparisons}
\end{table*}

\subsection{RQ2: Comparing Baseline and PCapGen}
\label{subsec:similarity-results}

Privacy experts labeled similarity (on a scale of 1-5) between PCapGen captions and the baseline captions across 90 code samples, and the average similarity scores for each of the PCapGen configurations are reported in Table \ref{tab:common-sample-similarity-ratings}. For comparison, we also report the average LLM-judge similarity scores for these 90 samples in the same table. We observe that all the similarity scores are $>2.4$ and most of them are $>3.3$, indicating that the PCapGen captions (across all configurations) are generally rated moderately similar to the developer-curated baseline captions by both privacy experts and LLM judges. 

When comparing the ratings by judges, we find that LlaMa and Gemini are slightly more conservative in their ratings (averaging 3.8 and 3.5, respectively) compared to Mistral (average rating of 4.2). This indicates that automatically generated PCapGen captions have moderate to high similarity with developer-curated baseline captions, and that they describe privacy behaviors but may differ in sentence structure, tone, or miss references to some privacy behaviors.
The second observation is that there is subjectivity in the assessment of caption similarity, especially among the privacy experts, despite aligning everyone on the similarity rubric before assessment. This can be observed by the difference in the similarity score ranges across each column in Table~\ref{tab:common-sample-similarity-ratings}. To understand this better, we qualitatively and quantitatively explore the agreement between experts and LLM judges. We initially compute complete and off-by-one (denotes the number of samples where the similarity rating of the sample differed exactly by 1 point) agreement counts between privacy experts and judges in Table \ref{tab:similarity-agreement-count}. 

\begin{table}[H]
\centering
\caption{Complete and off-by-one agreement counts between experts and judges.}
\begin{tabular}{|l|r|r|}
\hline
\textbf{Pair} & \textbf{Complete} & \textbf{Off-by-one} \\ \hline
Expert A vs Expert B     & 23 & 26 \\ \hline
Gemini vs Expert A        & 28 & 31 \\ \hline
Gemini vs Expert B        & 30 & 37 \\ \hline
Mistral vs Expert A       & 11 & 27 \\ \hline
Mistral vs Expert B       & 28 & 37 \\ \hline
LlaMa vs Expert A         & 16 & 36 \\ \hline
LlaMa vs Expert B         & 35 & 34 \\ \hline
\end{tabular}

\label{tab:similarity-agreement-count}
\end{table}

The privacy experts either completely agree or differ by 1 similarity rating on 49 out of 90 samples. Kirpendorff's Alpha inter-annotator agreement between them was only 0.168, which is considered low agreement. When we analyzed the differences in individual sample ratings between experts, we found that Expert A rates similarity by focusing on both \textit{how} personal information is used and \textit{why} it is used, whereas Expert B focused more on \textit{how} personal information is used. Consider the following pair of captions:

\begin{tcolorbox}[colback=gray!10,colframe=black!30,boxrule=0.25pt, boxsep=1pt]
\textbf{Baseline}: \textit{This app uses your current location, person ID, password, and service credentials to authenticate and authorize access to its services, which are then used to provide personalized content and functionality.}
\end{tcolorbox}

\begin{tcolorbox}[colback=gray!10,colframe=black!30,boxrule=0.25pt, boxsep=1pt]
\textbf{PCapGen}: \textit{This app accesses your device's location using GPS and network providers to determine your current position, then uses this information to construct URLs for fetching location-specific content. It also incorporates your personal identifiers, such as person ID and password, to personalize the content being retrieved.}
\end{tcolorbox}

Expert A rated the similarity as 2 (Slightly Similar) because PCapGen caption describes the \textit{why} part (`fetching location-specific content`) and the other does not, whereas Expert B rated the similarity as 4 (Highly Similar) focusing more on the overall description of privacy behavior, i.e., location information is collected to provide location-based services/features.

We believe this low agreement among the experts isn't a one-instance issue specific to our two experts, but rather a systemic problem where prior works have also shown that it is challenging to have high agreement between privacy experts \cite{jain2023towards, li2022Understanding}. 

Note that the privacy experts' assessments are consistent across different PCapGen configurations, so the difference observed is not PCapGen configuration/model dependent.

We also compute inter-annotator agreement between the LLM judges and find Kirpendorff's Alpha to be 0.6, which is considered moderate agreement, indicating that the judges are mostly in agreement among themselves. We also analyze the agreement between privacy experts and judges and find that Expert B is more in agreement with the judges as compared to Expert A. For example, Table~\ref{tab:similarity-agreement-count} shows that Expert B completely agrees or is off by 1 for 69 out of 90 samples with LlaMa, whereas the same count for Expert A is 52 out of 90 samples. 

This high agreement between a privacy expert and the LLM judges indicates that LLM judges are a viable alternative to privacy experts for measuring the quality of automatically generated captions.

\begin{table}[htbp]
\footnotesize
\centering
\caption{Similarity rating for common 90 samples.}
\begin{tabular}{|p{2.4cm}|p{1cm}|p{1cm}|p{1cm}|p{1cm}|p{1cm}|}
\hline
\textbf{Configuration} & \textbf{ExpertA} & \textbf{ExpertB} & \textbf{Mistral} & \textbf{Gemini} & \textbf{LlaMa} \\ \hline
PCapGen\_Claude   & 2.466 & 3.333 & 4.180 & 3.555 & 3.827 \\ \hline
PCapGen\_DeepSeek & 2.433 & 3.566 & 4.301 & 3.601 & 3.824 \\ \hline
PCapGen\_GPT      & 2.400 & 3.433 & 4.256 & 3.538 & 3.800 \\ \hline

\end{tabular}
\label{tab:common-sample-similarity-ratings}
\end{table}

Lastly, we report the BERTScore, BARTScore, and ROUGE-L scores between PCapGen and baseline captions in Table \ref{tab:similarity-metrics}. Both semantic similarity metrics -BERTScore and BARTScore- describe the baseline and PCapGen captions as being very similar, which is in line with what privacy experts and judges say. The average BERTScore precision is 0.91, whereas the BARTScore on average is -3.080. \footnote{For BARTScore, the closer the value to 0, higher the similarity.}

Between the PCapGen configurations, we do not find significant differences in the scores.

ROUGE-L scores compare token similarity across the PCapGen and baseline  captions, and we notice an average precision of 0.33 -- indicating that only a third of tokens overlap between the two captions. High semantic similarity scores but low ROUGE-L scores indicate that baseline and PCapGen captions use different vocabulary, yet are semantically very similar.

In conclusion, we observe that automatically generated PCapGen captions are semantically very similar to the developer-curated baseline captions, and while privacy experts do not always agree among themselves, LLM judges are a viable alternative to privacy experts for measuring the quality of automatically generated captions.

\begin{table*}[htbp]
\centering
\caption{Similarity metrics comparing baseline with PCapGen privacy captions.}
\begin{tabular}{|l|r|r|r|r|r|r|}
\hline
\textbf{Model} & \textbf{BERT P} & \textbf{BERT R} & \textbf{BERT F1} & \textbf{BARTScore} & \textbf{ROUGE-L P} & \textbf{ROUGE-L R} \\ \hline
PCapGen\_Claude   & 0.906 & 0.920 & 0.912 & -3.080 & 0.315 & 0.358 \\ \hline
PCapGen\_DeepSeek & 0.919 & 0.921 & 0.920 & -3.042 & 0.369 & 0.365 \\ \hline
PCapGen\_GPT    & 0.906 & 0.918 & 0.912 & -3.120 & 0.300 & 0.360 \\ \hline

\end{tabular}
\label{tab:similarity-metrics}
\end{table*}

\subsection{RQ3: Comparing Captions with Judges}
\label{subsec:llm-results}

\begin{table}[H]
\centering
\caption{Paired comparison between baseline and PCapGen\_Claude captions across quality criteria}
\begin{tabular}{|l|r|r|r|}
\hline
\textbf{Dimension} & \textbf{Mean Difference (b–a)} & \textbf{$p_{holm}$} & \textbf{$Cohen's\_ d_z$} \\ \hline
Completeness & 0.467 & 0.000 & 0.899 \\ \hline
Accuracy     & 0.572 & 0.000 & 0.859 \\ \hline
Conciseness  & 0.350 & 0.000 & 0.848 \\ \hline
\end{tabular}

\label{tab:llm-dimension-pairwise-stats}
\end{table}

Table \ref{tab:llm-rating} reports the average ratings for conciseness, completeness, and accuracy by LLM judges for the captions generated using the best configuration of PCapGen (i.e., PCapGen\_Claude, see Section~\ref{subsec:config_comparison}) and the baseline captions. 
Across all three criteria, the PCapGen\_Claude captions consistently outperform the baseline captions according to all judges. 

We run a paired t-test between the two configurations. The results are shown in Table \ref{tab:llm-dimension-pairwise-stats}.
On average, PCapGen\_Claude captions show substantial (all $Cohens\_d_z$ values are $>0.8$, indicating a large effect size) and statistically significant ($p_{holm} < 0.001$) improvements relative to the baseline: +0.35 in conciseness, +0.47 in completeness, and +0.57 in accuracy. 

These findings indicate that LLM judges consistently perceive PCapGen captions as more concise, complete, and accurate than baseline captions.
Beyond these quality metrics, we observe that all LLM judges preferred PCapGen\_Claude captions over the baseline captions: Llama (442/442), Gemini (370/442), and Mistral (426/442).

\subsection{RQ4: Comparing Captions with Experts}
\label{subsec:human-results}

\begin{table}[htbp]
\centering
\caption{Average privacy expert ratings across PCapGen and Baseline captions.}
\label{tab:avg_ratings}
\begin{tabular}{|l|r|r|r|}
\hline
\textbf{Configuration} & \textbf{Conciseness} & \textbf{Completeness} & \textbf{Accuracy} \\ \hline
PCapGen\_Claude & 3.883 & 3.900 & 3.858 \\ \hline
Baseline        & 3.775 & 3.675 & 3.625 \\ \hline
\end{tabular}
\label{tab:expert-dimension-ratings}
\end{table}

As stated in Section~\ref{sec:expert_survey_method}, we survey privacy experts to rate and compare PCapGen\_Claude captions with baseline captions. Note that due to the nature of our survey, for each of the 120 code samples, we only obtain quality ratings for either PCapGen\_Claude caption or the baseline caption, and not both. 

\begin{figure}[H]
  \centering
  \begin{subfigure}[t]{0.28\textwidth}
    \centering
    \includegraphics[width=0.85\linewidth]{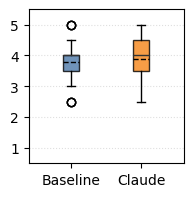}
    \caption{Conciseness}
    \label{fig:human-conciseness}
  \end{subfigure}
  \hfill
  \begin{subfigure}[t]{0.28\textwidth}
    \centering
    \includegraphics[width=0.85\linewidth]{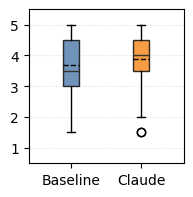}
    \caption{Completeness}
    \label{fig:human-completeness}
  \end{subfigure}
  \hfill
  \begin{subfigure}[t]{0.28\textwidth}
    \centering
    \includegraphics[width=0.85\linewidth]{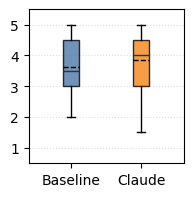}
    \caption{Accuracy}
    \label{fig:human-accuracy}
  \end{subfigure}
  \caption{Human evaluation ratings for PCapGen\_Claude and Baseline across three dimensions.}
  \label{fig:human-eval-ratings}
\end{figure}

Table \ref{tab:expert-dimension-ratings} summarizes the expert ratings of conciseness, completeness, and accuracy of both captions averaged over their 60 code samples.  Figures \ref{fig:human-conciseness}, \ref{fig:human-completeness}, and \ref{fig:human-accuracy} present whisker plots showcasing the rating distributions for the three criteria in Figure  \ref{fig:human-eval-ratings}. Across all three criteria, we observe that the median score for PCapGen\_Claude captions is higher than for the baseline captions, and very few PCapGen\_Claude samples score below 3.

Based on these trends, we observe that PCapGen\_Claude captions are of a higher quality compared to the baseline captions.
However, when comparing each criterion, we did not find any statistically significant difference ($p-value = 0.161$, $p-value = 0.166$, and $p-value =0.330$ for conciseness, completeness, and accuracy, respectively). Expert-provided quality metric annotations alone are sometimes not sufficient to demonstrate a statistically significant difference across configurations, which is also reported in a similar study \cite{su2024distilled}.  

To get a more conclusive result, we rely on the privacy expert provided preferences between PCapGen\_Claude and baseline captions (refer to the last part of the expert survey, that follows the design in Figure~\ref{fig:step-3}). We obtain caption preferences from each expert for all 120 samples.

Out of 240 comparisons (120 samples x 2 annotators) of both captions, PCapGen\_Claude was preferred 172/240 instances ($\sim$72\%). This result is similar to the one observed in Section \ref{subsec:llm-results}, where the judges also overwhelmingly preferred PCapGen\_Claude over the baseline. Overall, despite a lack of statistically significant difference in ratings of quality criteria, experts preferred automatically generated PCapGen\_Claude captions over the baseline captions.

\label{sec:results}

\section{Discussion}

In this section, we discuss the implications of our results, limitations of our method, and possible future directions. 

\textbf{Implications of the results.}
\label{subsec:implications}
Our results demonstrate that PCapGen captions outperform the baseline captions across multiple evaluation criteria and evaluators (i.e., LLMs-as-judge and privacy experts). This indicates that the taint analysis context enables LLMs to better understand and describe privacy behaviors in the code. Furthermore, the judges' preference for all PCapGen configurations over the baseline suggests that the performance is not strictly model dependent and is robust. When comparing the PCapGen configurations, PCapGen\_Claude's better performance suggests that certain models may be better suited for understanding privacy behaviors implemented in source code, which could be attributed to the training data or fine-tuning methods. 
We observe an interesting contrast in the results between the semantic similarity of PCapGen and baseline captions and the evaluators' preferences. Both metrics and evaluators indicate that PCapGen and baseline captions are highly similar semantically, but the quality criteria scores (especially for PCapGen\_Claude) and a strong preference for PCapGen captions show PCapGen is good at automatically generating captions that capture privacy nuances in the code. 
Furthermore, high semantic similarity and low n-gram score (ROUGE-L precision = 0.33) indicate that PCapGen rephrases privacy captions to be more concise, complete, and accurate than the baseline captions.

\textbf{Limitations.}
\label{subsec:limitations}
Our study has a few limitations that we attempt to mitigate to a large extent. A key limitation is the instrumentation in our study to generate privacy captions, i.e., both the LLM versions and the prompt. As some of the models we experiment with are proprietary and closed source, the model's behavior could change without notice. We report the exact model versions we used in this work to aid in reproducibility of the findings; however, the results can differ since we used a temperature of 0.7 -- which provides some room for variability in the LLM responses. Another aspect is the prompt used when interacting with LLMs. While we do careful prompt engineering to arrive at the final prompt version that gives good performance, the prompt space is too large to confirm if our final prompts are the best ones for each of our tasks. 

Another limitation is selection bias. Though we selected participants from Prolific's Domain Experts pool, we cannot guarantee their domain expertise. While programming-based questions were once a standard measure for screening participants, recent work has shown that these can be easily circumvented by AI \cite{ghorbani2023autonomy, prybylo2024evaluating}. To mitigate this limitation, we closely examined participants' responses and their completion times. If the completion time was exceptionally fast or the responses were inaccurate, we discarded the response and removed the participant from the study.

Lastly, our approach can also be impacted by false positives and false negatives. 
The \textit{Identifier} component misses privacy behaviors implemented by APIs in categories such as camera and storage, which is a limitation of our work, and we plan to address this in the future. However, for the other API categories, our approach works with all known and unknown sinks, when identifying taint paths, so we rarely miss any valid privacy behaviors or have false negatives. Since we use API calls to identify privacy behaviors, we have a low probability of false positives too. 



When the \textit{Extractor} maps the Jimple statements, false negatives/positives could impact the resultant privacy captions. However, we address both cases by mapping one Jimple statement at a time for each method. We initially attempted to map all Jimple statements belonging to a method in one LLM API call, but found several false negative cases arising. Switching to mapping only one Jimple statement in each LLM API call significantly reduced false negatives. In case of a false positive, where LLM identifies a statement in the source code that is not a correct mapping of the statement, it would not drastically impact the privacy caption generation since there are other taint analysis statements from other methods and the complete source code context for the LLM (in the \textit{Generator} component) to create concise, complete, and accurate privacy captions. Lastly, when creating the privacy captions, LLMs may miss privacy behaviors implemented in the source code and fail to describe them in the caption, or hallucinate behaviors. We observed such patterns with prompt chaining, where false negatives could propagate; therefore, we simplified the prompt to reduce these possibilities. Hallucination is an inherent limitation of LLMs, and it is a current limitation of our approach; however, we attempt to mitigate it by engineering our prompts to limit these possibilities. 

\textbf{Future Directions.}
There are several possible future directions for this work. First, while we demonstrate that our approach can be used with open-source models where developers do not have to worry about sharing sensitive information with third parties, it still relies on large models that cannot easily run on consumer hardware. In the future, we plan to fine-tune small language models that can run on consumer hardware, thereby allowing developers to create captions on their own machines. 
Second, since we extract the source code context that implements privacy behaviors, apart from generating privacy captions, we can also fine-tune the models to create privacy labels, thereby covering all notice formats. 
Third, while our current taint analysis approach captures taint paths for location, network, WiFi, and account-related APIs, it does not extract the paths for other APIs, such as storage and camera. We plan to further refine this by using machine learning to predict additional methods as potential sinks for a given source method. This would remove the dependency on the call graphs to identify taint paths, and ensure it captures taint paths of all APIs.

\label{sec:discussion}

\section{Conclusions}

In this paper, we introduced \textit{PCapGen}, an automated approach that identifies relevant code context in Android apps, extracts them, and generates privacy captions describing \textit{what} personal information is used, \textit{how} it is used, and \textit{why}. We develop a novel heuristics-based taint analysis approach to extract taint paths between known sources and unknown sinks. Using this approach, we create a dataset of 442 samples of Android code. We then iteratively design a prompt to generate privacy captions  
using three proprietary and open-source models 
and conduct a comprehensive evaluation with LLMs-as-judge, human privacy experts, and automated metrics. When comparing PCapGen configurations, we found that PCapGen\_Claude performs better than other approaches for our three quality criteria. Between PCapGen\_Claude and the baseline approaches, LLMs-as-judge consistently rated PCapGen captions higher than baseline and preferred PCapGen over baseline in a large majority of pairwise choices. Privacy experts showed no statistically significant mean differences for each criterion, yet favored PCapGen in pairwise comparisons, suggesting a superior performance of PCapGen captions over the baseline approach. 
Our work demonstrates the feasibility of automatically generating good-quality privacy captions directly from the source code; thereby aiding resource-constrained developers in fulfilling their regulatory obligations. 


\label{sec:conclusions}

\section{Data Availability}
\label{sec:data-availability}

We release our prompts and scripts to generate privacy captions and evaluate them with LLMs on our GitHub repository. We also share our annotation tool used in the survey in a separate repository \footnote{\url{https://github.com/PERC-Lab/PCapGen}} \footnote{\url{https://github.com/PERC-Lab/PCapGen-Annotation-Tool}}

\section{Ethics Consideration}
\label{sec:ethics-consideration}

This research adheres to our university's ethical guidelines and was conducted with the approval of our Institutional Review Board (IRB). All participants agreed to a consent form that included information about the investigators, risks, benefits, compensation, and confidentiality. All participants were informed that their participation was voluntary and that they could withdraw from the study at any time. The consent form also included the contact information of the investigators and the IRB team. No participant contacted the investigators or the IRB about the study. 

\section{LLM Usage Considerations}
\label{sec:llm-usage}

We describe the usage and limitation of LLMs in our methodology in Sections \ref{sec:approach}, \ref{sec:dataset}, \ref{sec:prompt-engg-exp}, and \ref{sec:discussion}. While we used larger models at this point, we describe our plan to use smaller models in future in Section \ref{sec:discussion}. We did not provide proprietary or open-source LLM any copyrighted, confidential, or personal information. We used open-source models along with proprietary models in our methodology to balance performance of proprietary models with the accessibility and lower carbon footprint of using open-source models. We did not use LLM for writing this paper.

\bibliographystyle{unsrt}  
\bibliography{main}

\section{Appendix: Prompt to Map Jimple Statements}
\label{appendix:map-jimple-prompt}

<Instruction>
You are an expert Java developer with an understanding of Jimple statements. You are given a source code and a statement in jimple. Your task is to identify which line of code does the jimple state represents.  In your response, only provide the line from the source code that matches the jimple statement and **do not provide any other text or explanation**. Return None if there are no matches. Do not create code from your own only return from the given source code. 
\newline
</Instruction>
\newline
\newline
<Example>
\newline
Source Code: {$example\_source\_code$}\newline
Statement: {$example\_statement$}\newline
Response: '$v1\_1$ = String.valueOf($v3\_2$.getLatitude());'\newline
</Example>
\newline

<Task>
\newline
Source Code: '{$src\_code$}'\newline
Statement: '{statement}'\newline
</Task>

\section{Appendix: Baseline Prompt}
\label{appendix:baseline-prompt}

<Instruction>
You are given a sample of an Android code snippet in JSON format. The samples are represented as a dictionary, where the key is the name of the class file and the value is the source code in that file. Your task is to write a privacy caption that best describe what the personal information being used in the code sample, how is it being used, and why. Ensure the caption is concise, complete, and accurate. 
\newline
</Instruction>

Following is an example of a code sample and its privacy caption. 

<Example> 
$code\_snippet$: {$example\_code\_snippet$}\newline
$example\_caption$: {$example\_privacy\_caption$}\newline
</Example> 

Here is the code snippet for you to generate the privacy caption. \newline
<Task>
\newline
$code\_snippet$: {$code\_sample$}\newline
</Task>

\section{Appendix: PCapGen Prompt}
\label{appendix:pcapgen-prompt}

<Instruction>
You are given a sample of an Android code snippet in JSON format. The samples are represented as a dictionary, where the key is the name of the class file and the value is the source code in that file. Your task is to write a privacy caption that best describe what the personal information being used in the code sample, how is it being used, and why. Ensure the caption is concise, complete, and accurate. Please pay special attention to the highlighted lines when writing the caption. Please only respond with the privacy caption, do not include any other text or explanation.
</Instruction>

Following is an example of a code sample along with highlighted lines and its privacy caption. 
    
<Example> 
\newline
$code\_snippet$: {$example\_code\_snippet$}\newline
$highlighted\_lines$: {$example\_highlighted\_lines$}\newline
$example\_caption$: {$example\_privacy\_caption$}\newline
</Example> 

Here is the code snippet for you to generate the privacy caption

<Task>
\newline
$code\_snippet$: {$code\_sample$}\newline
$highlighted\_lines$: {$highlighted\_lines$}\newline
</Task>

\section{Appendix: Prompt to Rate Privacy Captions}
\label{appendix:first-eval-prompt}

<Instruction> 
You are tasked with rating privacy captions, which are short sentences describing what personal information is used, how it is used, and why in a code snippet. You are given a sample of an Android code snippet in JSON format. The samples are represented as a dictionary, where the key is the name of the class file and the value is the source code in that file. You are also given some highlighted lines for some of the files in the code snippet as they could contain the core of the privacy behavior. Please pay special attention to these highlighted lines when reviewing. 

Rate the privacy caption based on conciseness, completeness, and accuracy on a scale of 1 to 5. Use the following rubric for each dimension

Conciseness
1: Extremely Verbose: aption is excessively verbose with significant redundancy and unnecessary details. Contains multiple irrelevant elements that obscure the main privacy behavior. Difficult to quickly understand due to wordiness.\newline
...    
\newline
5: Perfectly Concise: Caption is optimally concise, using the minimum words necessary to fully convey the privacy behavior. Every word serves a purpose. No redundancy or filler. Clear and immediately understandable.

Completeness
1: Very Incomplete: Caption misses critical privacy behavior elements. Fails to identify what data is involved, how it's used, or why. Reader cannot understand the basic privacy implications from the caption alone. 
\newline
..
\newline
5: Fully Complete: Caption comprehensively covers all aspects of the privacy behavior: what data, how it's processed, why, under what conditions, and any exceptions or limitations. Nothing important is missing for full understanding.

Accuracy
1: Highly Inaccurate: Caption contains multiple significant errors or misrepresentations of the code's privacy behavior. Fundamentally mischaracterizes what the code does with data. Would seriously mislead readers.
\newline
...
\newline
5: Perfectly Accurate: Caption precisely and correctly describes every aspect of the privacy behavior in the code. No errors, mischaracterizations, or misleading statements. Completely faithful to what the code actually does.

Please respond in the following format without including any markdown syntax:

{{
    "$conciseness$": <Rating for conciseness>,\newline
    "$conciseness\_rationale$": <Explanation for your rating>,\newline
    "$completeness$": <Rating for completeness>,\newline
    "$conciseess\_rationale$": <Explanation for your rating>,\newline
    "$accuracy$": <Rating for accuracy>,\newline
    "$accuracy\_rationale$": <Explanation for your rating>\newline
}}

</Instruction>
\newline

<Task>\newline
$code\_snippet$: {$code\_sample$}\newline
$highlighted\_lines$: {$highlighted\_lines$}\newline
$privacy\_caption$: {$privacy\_caption$}\newline
</Task>

\section{Appendix: Prompt to Choose Privacy Captions}
\label{appendix: second-eval-prompt}

<Instruction> 
You are given a sample of an Android code snippet in JSON format. The samples are represented as a dictionary, where the key is the name of the class file and the value is the source code in that file. You are also given some highlighted lines for some of the files in the code snippet as they could contain the core of the privacy behavior. Please pay special attention to these highlighted lines when reviewing. Your task is to choose the better privacy caption. Use the ratings for completeness, conciseness, and accuracy for the second privacy caption and evaluate which privacy caption best describes the privacy behavior of the given source code snippet. 

Please respond in the following format without including any markdown syntax:

{{
    "$caption\_choice$": <The number representing the better caption>\newline
    "$choice\_rationale$": <Explanation for your choice> \newline
}}

</Instruction>

<Task>
\newline
$code\_snippet$: {$code\_sample$} \newline
$highlighted\_lines$: {$highlighted\_lines$}\newline
[1] First Caption: {$ground\_truth$}\newline
[2] Second Caption: {$privacy\_caption$}\newline
</Task>

\end{document}